\documentclass[conference]{IEEEtran}
\usepackage{times}
\usepackage{footnote}
\usepackage{verbatim}
\usepackage{graphicx}  

\usepackage{psfrag}    
\begin{document}

\title{Cooperation Between Stations in Wireless Networks}

\author{\authorblockN{Andrea G. Forte and Henning Schulzrinne}
\authorblockA{
Columbia University\\
Email: \{andreaf, hgs\}@cs.columbia.edu}
}


 


\maketitle

\begin{abstract}
In a wireless network, mobile nodes (MNs) repeatedly perform 
tasks such as layer 2 (L2) handoff, layer 3 (L3) handoff and  
authentication. These tasks are critical, particularly for real-time 
applications such as VoIP.
We propose a novel approach, namely Cooperative Roaming (CR), in which MNs 
can collaborate with each other and share useful information about the 
network in which they move.

We show how we can achieve seamless L2 and L3 handoffs regardless
of the authentication mechanism used and without any changes to either the
infrastructure or the protocol. In particular, we provide a working 
implementation of CR and show how, with CR, MNs can achieve a total L2+L3 
handoff time of less than 16\,ms in an open network and of about 21\,ms 
in an IEEE 802.11i network. 
We consider behaviors typical of IEEE 802.11 networks, although 
many of the concepts and problems addressed here apply to any kind of 
mobile network.
\end{abstract}


%
\IEEEpeerreviewmaketitle

\section{Introduction} \label{intro}
Enabling VoIP services in wireless networks presents many 
challenges, including QoS, terminal mobility and congestion 
control. In this 
paper we focus on IEEE 802.11 wireless networks and 
address issues introduced by terminal mobility.

In general, a handoff happens when an MN moves out of the 
range of one Access Point (AP) and enters the range of a new one.
We have two possible scenarios:

\begin{enumerate}
\item If the old AP and the new AP belong to the same subnet, the 
MN's IP address does not have to change at the new AP. The MN 
performs a L2 handoff.
\item If the old AP and the new AP belong to different subnets, 
the MN has to go through the normal L2 handoff procedure 
and also has to request a new IP address in the new 
subnet, that is, it has to perform a L3 handoff.
\end{enumerate}

\begin{figure}
\centering
\rotatebox{270}{\includegraphics[height=3.5in]{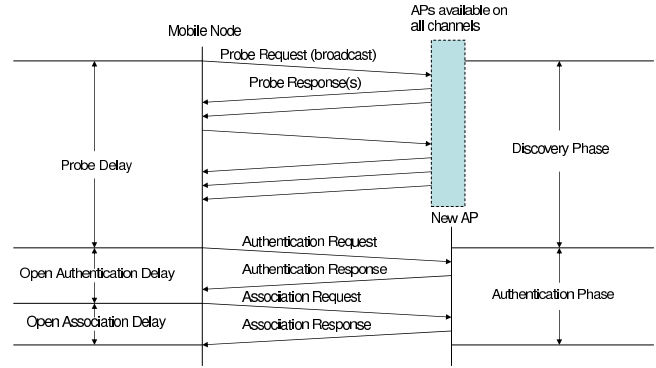}}
\caption{Layer 2 handoff procedure}
\label{L2proc}
\end{figure}

Fig. \ref{L2proc} shows the steps involved in a L2 handoff process in an open network. 
As we have shown in \cite{fastL2} and Mishra et al. have shown in \cite{L2analysis}, 
the time needed by an MN to 
perform a L2 handoff is usually on the order of a few hundred milliseconds, 
thus causing a noticeable interruption in any ongoing real-time multimedia 
session. In either open 802.11 networks or 802.11 networks with 
WEP enabled, the discovery phase constitutes more than 90\% of the
total handoff time\,\cite{fastL2, L2analysis}. In 802.11 networks with either WPA 
or 802.11i enabled, the handoff delay is dominated 
by the authentication process that is performed after associating to the 
new AP. In particular, no data can be exchanged amongst MNs before the  
authentication process completes successfully. 
In the most general case, both authentication delay and scanning delay
are present. These two delays are additive, so, in order 
to achieve seamless real-time multimedia sessions, 
both delays have to be addressed and, if possible, removed.

When a L3 handoff occurs, an MN has to perform a normal L2 handoff
and update its IP address.
We can break the L3 handoff into two logical steps: subnet change 
detection and new IP address acquisition via DHCP\,\cite{rfcDHCP}.
Each of these steps introduces a significant delay. 

In this paper we focus on the use of station cooperation
to achieve seamless L2 and L3 handoffs.
We refer to this specific use of cooperation as Cooperative Roaming (CR).
The basic idea behind CR is that MNs subscribe to the same multicast group creating a new plane for exchanging 
information about the network and help each other in different tasks. For example, an MN can discover 
surrounding APs and subnets by just asking to other MNs for this information. Similarly, an MN can ask 
another MN to acquire a new IP address on its behalf so that the first MN can get an IP address for the 
new subnet while still in the old subnet. 

For brevity and clarity's sake, in this paper we do not consider handoffs between different administrative 
domains and AAA-related issues although CR could be easily extended to support them. 
Incentives for cooperation are also not considered since they are a standard problem for any 
system using some form of cooperation (e.g., file sharing) and represent a separate research topic 
\cite{incentv1, incentv_game, incentv_econ, incentv2, incentv3}. 

The rest of the paper
is organized as follows. In Section \ref{rwork} we show the state
of the art for handoffs in wireless networks, in Section
\ref{multi} we briefly describe how IPv4 and IPv6 multicast addressing is used in the 
present context, Section \ref{cr} describes how,
with cooperation, MNs can achieve seamless L2 and L3 handoffs. Section \ref{auth} introduces cooperation 
in the L2 authentication process to achieve seamless handoffs regardless of the particular authentication mechanism used.
Section \ref{sec} considers security and Section \ref{sip} shows how streaming media can be supported in CR.
In Section \ref{energy} we analyze CR in terms of bandwidth and energy usage, 
Section \ref{exper} 
presents our experiments and results and 
Section \ref{SIPmob} shows how we can achieve 
seamless application layer mobility with CR. In Section \ref{loadb} 
we apply CR to load balancing and Section \ref{alternative} presents an alternative to multicast.
Finally, Section \ref{concl} concludes the paper.

\begin{table*}
\centering
\caption{L2 handoff time (ms)}
\begin{tabular}{|l|r|r|r|r|r|r|r|r|r|r|r|} \hline
Original Handoff & 457.8 & 236.8 & 434.8 & 317.0 & 566.7 & 321.6 & 241.0 & 364.0 & 216.7 & 273.9 & 343.0\\ \hline
Selective Scanning & 140.3 & 101.1 & 141.7 & 141.9 & 141.3 & 139.7 & 143.4 & 94.7 & 142.9 & 101.5 & 128.9\\ \hline
Cache & 2.7 & 2.4 & 4.2 & 3.7 & 4.4 & 2.6 & 2.6 & 2.3 & 2.7 & 2.9 & 3.0\\ \hline
\end{tabular}
\label{L2delay}
\end{table*}

\section{Related Work} \label{rwork}
The network community has done a lot of work on L2 
and L3 handoffs in wireless networks.
As of the writing of this paper, many standards such as IEEE 802.11f\,\cite{80211f} and
IEEE 802.11e\,\cite{80211e} have been ratified and others, such as IEEE 802.11k\,\cite{80211k_draft}, 
IEEE 802.11r\,\cite{80211r_draft} and IEEE 802.21\,\cite{80221_draft} are emerging, trying to solve 
some of the problems a wireless environment introduces.
All of these approaches, however, introduce significant changes in 
the infrastructure and in the protocol. In particular, they have 
always been structured thinking of each MN as a stand alone 
entity.

802.11f focuses on ways in which APs can 
share information among each other with the definition of an Inter Access 
Point Protocol (IAPP). This can be particularly useful for the transfer of users' credentials
during handoffs, for example.

The 802.11e protocol addresses QoS problems in wireless Local Area Networks (LANs). In particular,
different traffic classes are defined with their own medium access
parameters, giving real-time traffic higher priority in accessing 
the wireless medium than best-effort traffic.

The 802.11k protocol utilizes MNs to collect topology information 
and other useful statistics about the network and conveys it back to the APs. The APs then build
a neighbor report containing all the information about the various
APs and their neighbors. These reports are then sent to the MNs so
that each MN can have information about its neighboring APs. The way these
reports are built is not specified and often involves
each MN having to scan different channels.

No draft has been ratified by the 802.11r working group as yet. 802.11r
addresses the need for fast L2 roaming in 802.11 networks considering 
different authentication mechanisms as well as QoS.
In 802.11r, fast Basic Service Set (BSS) transitions can only take place 
between APs in the same mobility domain. A mobility 
domain is a set of BSSs in the same Extended Service Set (ESS). 
Within a mobility domain, APs can exchange key material and context using 
encapsulation over the distribution system.
802.11r does not specify how an MN discovers the best candidate AP to 
connect to next. Scanning, neighbor reports and other means can be used.
802.11r supports pre-keying and resource reservation between MN and AP and it 
defines a key hierarchy to extend Pairwise Master Keys (PMKs) to multiple APs.

The IEEE 802.21 (Media Independent Handover) standard \cite{80221_draft} introduces link-layer enhancements 
for performing intelligent handoffs between heterogeneous networks such as IEEE 802.11 and 
cellular, including both wireless and wired networks. The handoff process can be initiated by either 
the client or the network, and just like in IEEE 802.11k,
MNs provide information about available networks and 
other network statistics to the infrastructure by scanning. The infrastructure then builds and stores information such as 
neighborhood cell lists and available services, thus helping in the optimum cell 
selection. Furthermore, new link-layer primitives are defined in order to provide applications 
with consistent information regardless of the access technology used by the MN.

In all these approaches, MNs always behave as stand alone entities often 
having to scan the medium before handoffs, that is, causing interruptions in any ongoing 
multimedia session. Furthermore, seamless 
handoffs with these approaches, when possible, require changes in the network and in the 
clients. CR is a client-only approach and can represent either an alternative 
or a complement to the current standards.

More recently, cooperative approaches have been proposed in the 
network community.
Liu et al.\,\cite{coopMAC} show how cooperation amongst 
MNs can be beneficial for all 
the MNs in the network in terms of bit-rate, coverage and throughput.
Each MN builds a table in which possible helpers for
that MN are listed. If an MN has a poor link with the AP and its
bit-rate is low, it sends packets to the helper who relays
them to the AP. The advantage in doing this is that the link from the MN to 
the helper and from the helper to the AP is a high bit-rate link. In 
this way the MN can use two high bit-rate links via the helper instead 
of the low bit-rate one directly to the AP.

Fretzagias et al.\,\cite{coopLoc} introduce a location sensing mechanism 
based on cooperative behavior among stations. Stations share 
location information about other stations and about landmarks so 
to improve position prediction and reduce training.

Other work uses a cooperative approach but mostly 
in positioning applications\,\cite{coopRobot}, sensor networks\,\cite{ButtyanHS05} 
and at the physical\,\cite{phyCoop, coopDiv} and application\,\cite{7ds} layers.

Aside from cooperation approaches and standardization efforts in the IEEE 802.11 working
groups, many other approaches have been proposed in order to achieve
fast handoffs in wireless networks. However, most of these approaches, 
such as \cite{smip} and \cite{sipbased}, require changes to either the infrastructure
or the protocol or both. One good example of such a situation is Mobile IP (MIP).
MIP has been standardized for many years now, however, it has never had a 
significant deployment, in part because of the considerable changes required 
in the infrastructure. Fast handoff approaches in the MIP context usually
require additional network elements\,\cite{MIPhw, wu-bidir} and/or changes 
to the protocol\,\cite{ro-mip}. 

In \cite{syncscan} Ramani et al. suggest an algorithm called syncscan
which does not require changes to either the protocol or the infrastructure. 
It does require, however, that all the APs in the network are 
synchronized and only accelerates unauthenticated L2 handoffs. 

In this paper we propose a novel approach that works in an already 
deployed wireless environment, an environment with heterogeneous
networks, where new network elements cannot necessarily be introduced in the infrastructure,
where all the APs are not necessarily synchronized amongst themselves,
where any kind of authentication mechanism can be used and where different
subnets may be present.

We use a cooperative approach amongst MNs 
for spreading information regarding the network topology without any
infrastructure support. Our approach requires changes
only to the wireless card driver, DHCP client and authentication 
supplicant; no changes to the infrastructure or the protocol are required.
This allows us to solve many of the problems associated with 
terminal mobility, regardless of the network the user
moves to.

\section{IP Multicast Addressing} \label{multi}
CR works for both IPv4 and IPv6. In IPv4, we make extensive use of 
UDP-over-IP multicast packets.
Different values for time-to-live (TTL) are used according to how far we want 
multicast packets to reach into the IP network. This also depends on the density of
MNs supporting the protocol. For example, if an MN does not receive any
response after sending a request with a TTL value of 1
(same subnet), it will send the same request again but with a TTL value
of 2 (next subnet) and so on. 
We must note, however, that the probability for an MN to find the 
information it needs becomes smaller as the search moves to 
more distant subnets.
On the other hand, a small TTL can be used to limit the propagation 
of CR multicast frames in very congested environments.

In IPv6, we would use multicast scopes instead of IPv4 multicast. No 
significant changes would be required.

\section{Cooperative Roaming} \label{cr}
In this section we show how MNs can cooperate with each other
in order to achieve seamless L2 and L3 handoffs.

\subsection{Overview} \label{L2-roam-ovrvw}
In \cite{fastL2} we have introduced a fast MAC layer handoff
mechanism for achieving seamless L2 handoffs in environments such
as hospitals, schools, campuses, enterprises, and other places where MNs 
always encounter the same APs.
Each MN saves information regarding the surrounding APs 
in a cache. When an MN needs to perform a handoff and
it has valid entries in its cache, it will directly use the information 
in the cache without scanning. If it does not have any valid
information in its cache, the MN will use an optimized scanning procedure 
called \textit{selective scanning} to discover new APs and build the cache.
In the cache, APs are ordered according to their signal strength that was registered
when the scanning was performed, that is, right before changing AP. APs with stronger signal 
strength appear first. 
As mentioned in Section \ref{intro}, in open networks the scanning process is responsible 
for more than 90\% of the total handoff time. 
The cache reduces the L2 handoff time 
to only a few milliseconds (see Table \ref{L2delay}) and 
cache misses due to errors in movement prediction introduce 
only a few milliseconds of additional delay\,\cite{fastL2}. Such an approach, however, works only
in open networks or networks with WEP enabled. Other forms of authentication
are not supported.

Earlier, we had extended \cite{fastL3} the mechanism introduced in \cite{fastL2}
to support L3 handoffs. MNs also cache 
L3 information such as their own IP address, default router's IP address and subnet identifier. A subnet 
identifier uniquely identifies a subnet. By caching the 
subnet identifier, a subnet change is detected much faster and L3 handoffs are triggered 
every time the new AP and old AP have 
different subnet identifiers. Faster 
L3 handoffs can be achieved since IP address and default router for the next AP and subnet 
are already known and can be immediately used. The approach in \cite{fastL3} achieves seamless handoffs in 
open networks only, it utilizes the default router's IP address as subnet identifier and it uses a 
suboptimal algorithm to acquire L3 information. 

\begin{figure}
\centering
\begin{tabular}{|l|r|r|r|} \hline
 & Current AP & Next Best AP & Second Best AP\\ \hline
BSSID & MAC A & MAC B & MAC C\\ \hline
Channel & 6 & 11 & 1\\ \hline
Subnet ID & 160.39.5.0 & 160.39.10.0 & 160.39.10.0\\ \hline
\end{tabular}
\caption{Example of MN's cache structure}
\label{cacheStruct}
\end{figure}

Here, we consider the same caching mechanism
used in \cite{fastL3}. In order to support multi-homed routers, however, we use
the subnet address as subnet identifier. By knowing the subnet mask and the default router's IP 
address we can calculate the network address of a certain subnet. Fig. \ref{cacheStruct} shows the 
structure of the cache. Additional information such as last IP address used by the MN, lease
expiration time and default router's IP address can be extracted from the
DHCP client lease file, available in each MN.

In CR, an MN needs to acquire information about the 
network if it does not have any valid information in the 
cache or if it does not have L3 information available for a 
particular subnet. 
In such a case, the MN asks other MNs for 
the information it needs so that the 
MN does not have to find out about neighboring APs 
by scanning. In order to share information, in CR, all MNs subscribe to 
the same multicast group.
We call an MN that needs to acquire information about its 
neighboring APs and subnets a requesting MN (R-MN). By using CR, an R-MN can ask 
other MNs if they have such information by sending an 
INFOREQ multicast frame. The MNs that receive 
such a frame check if they have the information the R-MN needs 
and if so, they send an INFORESP multicast frame back to the R-MN
containing the information the R-MN needs. 

\subsection{L2 Cooperation Protocol} \label{L2-roam}
In this section, we focus on the information exchange needed by a L2 
handoff.

The information exchanged in the INFOREQ and INFORESP frames is a list of
\{BSSID, channel, subnet ID\} entries, one for each AP in the 
MN's cache (see Fig. \ref{cacheStruct}). 

When an R-MN needs information about its neighboring APs and subnets, 
it sends an INFOREQ multicast frame. Such a frame 
contains the current content of the R-MN's cache, that is, all APs and 
subnets known to the R-MN. 
When an MN receives an INFOREQ frame, it checks if 
its own cache and the R-MN's cache have at least one AP in common. 
If the two caches have at least one AP in common and if the MN's cache has 
some APs that are not present in the R-MN's 
cache, the MN sends an INFORESP multicast frame
containing the cache entries for the missing APs. 
MNs that have APs in common with the R-MN, have been in the same 
location of the R-MN and so have a higher probability of having the 
information the R-MN is looking for. 

The MN sends the INFORESP frame after waiting for a random amount of time to be 
sure that no other MNs have already sent such information.
In particular, the MN checks the information contained in
INFORESP frames sent to the same R-MN by other MNs during
the random waiting time.
This prevents many MNs from sending the same information 
to the R-MN and all at the same time.

When an MN other than R-MN receives an INFORESP 
multicast frame, it 
performs two tasks. First, it checks if someone is 
lying by providing the wrong information and if so, it tries to fix 
it (see Section \ref{roam-sec}); secondly, it records the 
cache information provided by such a frame in its cache even though
the MN did not request such information.
By collecting unsolicited information, each MN can build a 
bigger cache in less time and in a more efficient manner requiring 
fewer frame exchanges. This is very similar to what happens in software 
such as Bit-Torrent where the client downloads different parts of the 
file from different peers. Here, we collect different cache chunks 
from different MNs.

In order to improve efficiency and further minimize frame exchange, 
MNs can also decide to collect information contained in the INFOREQ frames.

\subsection{L3 Cooperation Protocol} \label{L3-roam}
In a L3 handoff an MN has to detect a change in subnet and also has to acquire a 
new IP address. When a L2 handoff occurs, the MN compares the cached subnet 
identifiers for the old and new AP. If the two identifiers are 
different, then the subnet has changed. 
When a change in subnet is detected, the MN needs to acquire a new 
IP address for the new subnet. The new IP address is usually acquired 
by using the DHCP infrastructure. Unfortunately, the 
typical DHCP procedure can take up to one second\,\cite{fastL3}. 

CR can help MNs acquire a new IP address for the new subnet 
while still in the old subnet.
When an R-MN needs to perform a L3 handoff, 
it needs to find out which other MNs in the new subnet 
can help. We call such MNs Assisting MNs 
(A-MNs).
Once the R-MN knows the A-MNs for the new subnet, it 
asks one of them to acquire a new IP address on its behalf. 
At this point, the selected A-MN acquires the new IP address via DHCP and sends it 
to the R-MN which is then able to update its 
multimedia session before the actual L2 handoff and can start using the new IP address right after the L2 handoff, hence 
not incurring any additional delay (see Section \ref{SIPmob}).

We now show how A-MNs can be 
discovered and explain in detail how they can request 
an IP address on behalf of other MNs in a different subnet.

\subsubsection{A-MNs Discovery} \label{discov} 
By using IP multicast, 
an MN can directly talk to different MNs in different subnets.
In particular, the R-MN sends an AMN\_DISCOVER 
multicast packet containing the new subnet ID. 
Other MNs receiving such a packet check the subnet 
ID to see if they are in the subnet specified in the 
AMN\_DISCOVER. If so, they reply with an AMN\_RESP 
unicast packet. This packet contains the A-MN's default router 
IP address, the A-MN's MAC and IP addresses. This 
information is then used by the R-MN to build a list of 
available A-MNs for that particular subnet.

Once the MN knows which A-MNs are available in the new subnet, it 
can cooperate with them in order to acquire the L3 information it needs 
(e.g., new IP address, router information), as described 
below.

\subsubsection{Address Acquisition} \label{addracq}
When an R-MN needs to acquire a new IP address for a particular 
subnet, it sends a unicast IP\_REQ packet to one of the 
available A-MNs for that subnet. Such packet contains the 
R-MN's MAC address. When an A-MN receives an IP\_REQ packet, 
it extracts the R-MN's MAC address from the packet and 
starts the DHCP process by inserting the R-MN's MAC address in the 
CHaddr field of DHCP packets\footnote{If supported, the client-ID field must 
be used instead\,\cite{rfcClientID}.}. The A-MN will also have to set 
the broadcast bit in the DHCP packets in order for it to receive DHCP packets with a 
different MAC address other than its own in the CHaddr field.
All of this allows the A-MN to acquire a new IP address on behalf 
of the R-MN. This procedure is completely transparent to the 
DHCP server.
Once the DHCP process has been completed, the A-MN sends an 
IP\_RESP multicast packet containing the default router's IP address 
for the new subnet, the R-MN's MAC address and the new IP address 
for the R-MN.
The R-MN checks the MAC address in the IP\_RESP packet to be 
sure that the packet is not for a different R-MN. 
Once it has verified that the IP\_RESP is for itself, the 
R-MN saves the new IP address together with the new default router's 
IP address.

If the R-MN has more than one possible subnet to move to, it 
follows the same procedure for each subnet. 
In this way the R-MN builds a list of \{router, new IP address\} 
pairs, one pair for each one of the possible next subnets. After 
moving to the new subnet the R-MN renews the lease for the new 
IP address.
The R-MN can start this process at any time before the L2 handoff, 
keeping in mind that the whole process might take one second or more 
to complete and that lease times of IP addresses are usually on the 
order of tens of minutes or more\footnote{The DHCP client lease file can 
provide information on current lease times.}. 

By reserving IP addresses 
before moving to the new subnet, we could waste IP 
addresses and exhaust the available IP pool.
Usually, however, the lease time in a mobile environment 
is short enough to guarantee a sufficient re-use of IP addresses.

Acquiring an IP address from a different subnet other than the 
one the IP is for could also be achieved by introducing a new 
DHCP option. Using this option, the MN could ask the DHCP server 
for an IP address for a specific subnet. This would however, 
require changes to the DHCP protocol. 

\section{Cooperative Authentication} \label{auth}
In this section we propose a cooperative approach for authentication
in wireless networks. The proposed approach is
independent of the particular authentication mechanism used. It
can be used for VPN, IPsec, 802.1x or any other kind of authentication. We 
focus on the 802.1x framework used in Wi-Fi Protected 
Access (WPA) and IEEE 802.11i\,\cite{80211i}.

\subsection{IEEE 802.1x Overview}
The IEEE 802.1x standard defines a way to perform access control
and authentication in IEEE 802 LANs and in particular in IEEE 
802.11 wireless LANs using three main entities:
supplicant, authenticator and authentication server\footnote{The authentication
server is not required in all authentication mechanisms.}. The supplicant
is the client that has to perform the authentication in order to gain access to the
network; the authenticator, among other things, relays packets between supplicant and 
authentication server; the authentication server, typically a RADIUS 
server\,\cite{rfcRadius}, performs the authentication process with the supplicant by
exchanging and validating the supplicant's credentials. 
The critical point, in terms of handoff time in the 802.1x architecture, is that 
during the 
authentication process the authenticator allows only EAP Over LAN (EAPOL) traffic
to be exchanged with the supplicant. No other kind of traffic is
allowed.

\begin{figure}
\centering
\rotatebox{270}{\includegraphics[height=3.5in]{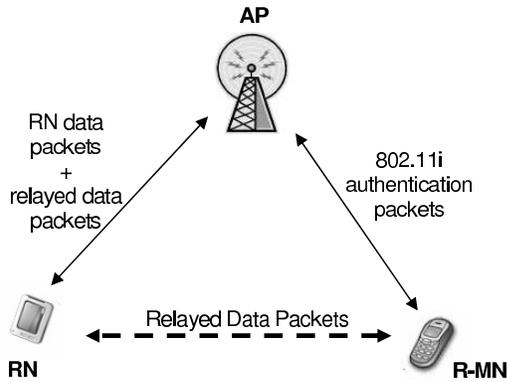}}
\caption{Layer 2 handoff with authentication in CR}
\label{relay}
\end{figure}

\subsection{Cooperation in the Authentication Process} \label{auth-relay}
A well-known property of the wireless medium in IEEE 802.11 networks
is that the medium is shared and therefore every MN can hear
packets that other stations (STAs) send and receive. This
is true when MN and STAs are connected to the same AP - that is, are on the 
same channel. In \cite{coopMAC} Liu et al. make use
of this particular characteristic and show how MNs can cooperate with
each other by relaying each other's packets so to achieve
the optimum bit-rate. In this section we show how a similar
approach can be used for authentication purposes.

For simplicity, in the following discussion we suppose that one 
authenticator manages one whole subnet, so that authentication is required
after each L3 handoff. In such a scenario and in this context, we also 
refer to a subnet as an Authentication Domain (AD).
In general, an MN can share the information about ADs in
the same way it shares information about subnets. In doing so, an MN 
knows whether the next AP belongs to the same AD of the current AP 
or not.
In a L2 or L3 handoff we have an MN which performs handoff and authentication, 
a Correspondent Node (CN) which has an established multimedia 
session with the MN and a Relay Node (RN) which relays 
packets to and from the MN. Available RNs for a particular AD can be discovered
following a similar procedure to the one described earlier for the discovery 
of A-MNs (see Section \ref{discov}). The difference here is that RN and MN have
to be connected to the same AP after the handoff. In this scenario, we assume 
that RNs are a subset of the available A-MNs. 
The basic idea is that while the MN is authenticating in the new AD,
it can still communicate with the CN via the RN which relays 
packets to and from the MN (see Fig. \ref{relay}).
Let us look at this mechanism in more detail. 
Before the MN changes AD/AP, it selects an RN from the list of 
available RNs for the new AD/AP and sends a RELAY\_REQ multicast frame 
to the multicast group. The RELAY\_REQ frame 
contains the MN's MAC and IP addresses, the 
CN's IP address and the selected RN's MAC 
and IP addresses.
The RELAY\_REQ will be received by all the STAs subscribed to the
multicast group and, in particular, 
it will be received by both the CN\footnote{In congested environments, where smaller TTL values 
may be preferred, a separate unicast RELAY\_REQ frame can be sent to the CN.} and the RN. The RN will relay 
packets for the MN identified by the MAC address received in the RELAY\_REQ frame. 
After performing the handoff, the MN needs to authenticate before
it can resume any communication via the AP. However, because of the
shared nature of the medium, the MN will start sending packets to the RN as if 
it was already authenticated. The authenticator will drop the packets,
but the RN can hear the packets on the medium and relay them
to the CN using its own encryption keys, that is, using its secure connection 
with the AP. The CN is aware of the relaying because of the RELAY\_REQ, 
and so it will start sending packets for the MN to the RN as well.
While the RN is relaying packets to and from the MN, the MN will perform
its authentication via 802.1x or any other mechanism. Once the authentication
process is over and the MN has access to the infrastructure, it can stop
the relaying and resume normal communication via the AP. When this
happens and the CN starts receiving packets from the MN via the AP, it
will stop sending packets to the RN and will resume normal communication
with the MN. The RN will detect that it does not need to relay any
packet for the MN any longer and will return to normal operation.
 
In order for this relaying mechanism to work with WPA and 802.11i, 
MN and RN have to exchange unencrypted L2 data 
packets for the duration of the relay process. These packets are 
then encrypted by the RN by using its own encryption keys and are sent to
the AP.
By responding to an RN discovery, RNs implicitly agree to providing
relay for such frames. Such an exchange of unencrypted L2 frames does not
represent a security concern since packets can still be encrypted at 
higher layers and since the relaying happens for a very limited amount 
of time (see Section \ref{sec-auth}). 

One last thing worth mentioning is that by using a relay, we remove the 
bridging delay in the L2 handoff\,\cite{fastL2, L2analysis}. Usually, after an MN changes AP, the switch continues
sending packets for the MN to the old AP until it updates the information regarding
the new AP on its ports. The bridging delay is the amount of time
needed by the switch to update this information on its ports. When we use a relay node in the new AP,
this relay node is already registered to the correct port on the switch, therefore no update
is required on the switch side and the MN can immediately receive packets via the RN.

\subsection{Relay Process} \label{relay-proc}
In the previous section we have shown how an MN can perform authentication while having 
data packets relayed by the RN. In this section we explain in more detail how relaying is performed.

\begin{figure}
\centering
\rotatebox{270}{\includegraphics[height=3.5in]{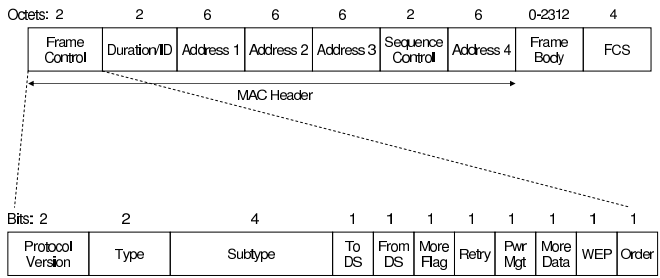}}
\caption{General IEEE 802.11 MAC layer frame format}
\label{mac_frame}
\end{figure}

Fig. \ref{mac_frame} shows the format of a general IEEE 802.11 MAC layer frame. Among the many fields we can identify a \textit{Frame Control} field 
and four \textit{Address} fields. For the relay process we are interested in the four \textit{Address} fields 
and in the \textit{To DS} and \textit{From DS} one-bit fields that are part of the \textit{Frame Control} field.
The \textit{To DS} bit is set to one in data frames that are sent to the Distribution System (DS)\footnote{A DS 
is a system that interconnects BSSs and LANs to create an ESS\,\cite{80211std}.}. 
The \textit{From DS} bit is set to one in data frames exiting the DS. 
The four \textit{Address} fields have a different meaning according to the particular combination 
of the \textit{To DS} and \textit{From DS} bits. Table \ref{addr_flds} shows the different meanings 
of the \textit{Address} fields for each combination of the \textit{To DS} and \textit{From DS} bits. The 
elements appearing in Table \ref{addr_flds} are: Destination Address (DA), Source Address (SA), BSSID, Receiver Address (RA) and 
Transmitter Address (TA).

\begin{table}
\centering
\caption{IEEE 802.11 MAC layer frame: Address fields content}
\begin{tabular}{|c|c|c|c|c|c|} \hline
To DS & From DS & Address 1 & Address 2 & Address 3 & Address 4\\ \hline
0 & 0 & DA & SA & BSSID & N/A\\ \hline
0 & 1 & DA & BSSID & SA & N/A\\ \hline
1 & 0 & BSSID & SA & DA & N/A\\ \hline
1 & 1 & RA & TA & DA & SA\\ \hline
\end{tabular}
\label{addr_flds}
\end{table}

In infrastructure mode, when an MN sends a packet, this packet is always sent first to the AP even if both source 
and destination are associated with the same AP. For such packets the MN sets the \textit{To DS} bit. 
Other MNs on the same channel can hear the packet but discard it because, as the \textit{To DS} field and 
\textit{Address} fields suggest, such 
packet is meant for the AP.
When the AP has to send a packet to an MN, it sets the \textit{From DS} bit. All MNs that can hear this packet 
discard it, except for the MN the packet is for.

When both fields, \textit{To DS} and \textit{From DS}, have a value of one, the packet is sent on the wireless medium 
from one AP to another AP. 
In ad-hoc mode, both fields have a value of zero and the frames are directly exchanged between MNs with the same 
Independent Basic Service Set (IBSS).

In \cite{v-wifi} Chandra et al. present an optimal way to continuously switch 
a wireless card between two or more infrastructure networks or between infrastructure and ad-hoc networks 
so that the user has the perception of being connected to multiple networks at the same time although using one 
single wireless card. This approach works well if no real-time traffic is present.
When we consider real-time traffic and its delay constraints, continuous switching between different 
networks and, in particular, between infrastructure and ad-hoc mode is no longer a feasible solution. 
Although optimal algorithms have been proposed for this\,\cite{v-wifi}, the continuous switching of the channel 
and or operating mode takes a non-negligible amount of time which becomes particularly significant if any form 
of L2 authentication is present in the network. In such cases, the time needed by the wireless card to continuously 
switch between networks can introduce significant delay and packet loss.

The approach we propose is based on the idea that ad-hoc mode and infrastructure mode do not have to 
be mutually exclusive, but rather can complement each other. In particular, MNs can send ad-hoc packets 
while in infrastructure mode so that other MNs on the shared medium, that is, on the same channel, can receive such 
packets without involving the AP. Such packets 
use the 802.11 ad-hoc MAC addresses as specified in \cite{80211std}. That is, both fields \textit{To DS} and 
\textit{From DS} have a value of zero and the \textit{Address} fields are set accordingly as specified in Table
\ref{addr_flds}. 
In doing so, MNs can directly send and receive packets to and from other MNs without involving the AP and without 
having to switch to ad-hoc mode. 

This mechanism allows an RN to relay packets to and from an R-MN without significantly affecting any ongoing multimedia 
session that the RN might have via the AP.
Such an approach can be useful in all those scenarios where an MN in infrastructure mode needs to communicate 
with other MNs in infrastructure or ad-hoc mode\,\cite{wifi-profiler} and a continuous change between infrastructure 
mode and ad-hoc mode is either not possible or convenient.

\section{Security} \label{sec}
Security is a major concern 
in wireless environments. In this section we 
address some of the problems encountered in a cooperative
environment, focusing on CR.

\subsection{Roaming Security Issues} \label{roam-sec}
In this particular context, a malicious user might try 
to propagate false information among the cooperating 
MNs. In particular, we have to worry about three 
main vulnerabilities:

\begin{enumerate}
\item A malicious user might want to re-direct STAs to 
fake APs where their traffic can be sniffed and private
information can be compromised.
\item A malicious user might try to perform DoS attacks
by re-directing STAs to far or non-existing APs. This would
cause the STAs to fail the association to the next AP during
the handoff process. The STA would then have to rely on the legacy 
scanning process to re-establish network connectivity.
\item At L3, a malicious user might behave as an A-MN and
try to disrupt a STA' service by providing invalid IP addresses.
\end{enumerate}

In general, we have to remember that the cooperative mechanism
described here works on top of any other security mechanism that
has been deployed in the wireless network (e.g., 802.11i, WPA). 
In order for a malicious user to send and receive packets from
and to the multicast group, it has to have, first of all, access 
to the network and thus be authenticated. 
In such a scenario, a malicious user is a STA with legal
access to the network. This means that MAC spoofing attacks
are not possible as a change in MAC address would require 
a new authentication handshake with the network. This also means that once 
the malicious user has been identified, it can be isolated.

How can we attempt to isolate a malicious node? 
Since the INFORESP frame is multicast, each 
MN that has the same information than the one contained in 
such a frame, can check that the information in such a 
frame is correct and that no one is lying. If it finds out 
that the INFORESP frame contains the wrong information, 
it immediately sends an INFOALERT multicast frame. Such a
frame contains the MAC address of the suspicious STA. This
frame is also sent by an R-MN that has received a wrong IP address 
and contains the MAC address of the A-MN that provided that 
IP address.
If \textit{more than one} alert for the \textit{same} suspicious node, 
is triggered by \textit{different} nodes, the suspicious 
node is considered malicious and the information it provides is ignored. 
Let us look at this last point in more detail.

One single INFOALERT does
not trigger anything. In order for an MN to be categorized as bad, there has to be a certain number of
INFOALERT multicast frames sent by \textit{different} nodes, all regarding the \textit{same} 
suspicious MN. This certain number can be configured according
to how paranoid someone is about security but, regardless, it has to be more than one. Let us assume this number to be five.
If a node receives five INFOALERT multicast frames from five different nodes regarding the same MN, then it marks 
such an MN as bad. This mechanism could be compromised if either a
malicious user can spoof five different MAC addresses (and this is not likely for the reasons we have explained earlier) or
if there are five different malicious users that are correctly authenticated in the wireless network 
and that can coordinate their attacks.
If this last situation occurs, then there are bigger problems in the network to worry about than handoff policies.
Choosing the number of INFOALERT frames required to mark a node as malicious to be very large would have 
advantages and disadvantages. It would give more
protection against the exploitation of this mechanism for DoS attacks as the number of malicious users 
trying to exploit INFOALERT frames would have to be high. On the other hand, it would also make
the mechanism less sensitive to detect a malicious node as the number of INFOALERT frames
required to mark the node as bad might never be reached or it might take too long to reach. So, there 
is clearly a trade-off. 

Regardless, in either one of the three situations described at the beginning of this section, 
the MN targeted by the malicious user would be able to easily 
recover from an attack by using legacy mechanisms such as 
active scanning and DHCP address acquisition, typically 
used in non-cooperative environments.

\subsection{Cooperative Authentication and Security} \label{sec-auth}
In order to improve security in the relay process, we introduce
some countermeasures that nodes can use to prevent exploitation
of the relay mechanism.
The main concern in having a STA relay packets for an unauthenticated
MN is that such an MN might try to repeatedly use the relay mechanism and 
never authenticate to the network. In order to prevent this, we
introduce the following countermeasures:
\begin{enumerate}
\item Each RELAY\_REQ frame allows an RN to relay packets for
a limited amount of time. After this time has passed, the relaying
stops. The relaying of packets is required
only for the time needed by the MN to perform the normal authentication
process.
\item An RN relays packets only for those nodes which have sent a
RELAY\_REQ packet to it while still connected to their previous AP.
\item RELAY\_REQ packets are multicast. All the nodes in the
multicast group can help in detecting bad behaviors such as one
node repeatedly sending RELAY\_REQ frames. 
\end{enumerate}
All of the above countermeasures work if we can be sure of the identity of a 
node and, in general, this is not always the case as malicious users can perform MAC 
spoofing attacks, for example. However, as we have explained in Section \ref{roam-sec}, 
MAC spoofing attacks are not possible in the present framework. 

This said, we have to remember that before an RN can relay packets for an MN, 
it has to receive the proper RELAY\_REQ packet from the MN. Such a 
packet has to be sent by the MN while still connected to the old AP. This means
that the MN has to be authenticated with the previous AP in order to send
such a packet. Furthermore, once the relaying timeout has expired, the 
RN will stop relaying packets for that MN. At this point, even if the MN can change its 
MAC address, it would not be able to send a new RELAY\_REQ as it has
to first authenticate again with the network (e.g., using 802.11i) and therefore 
no relaying would take place.
In the special case in which the old AP belongs to an open network\footnote{Under normal 
conditions this is very unluckily but it might happen for handoffs between different 
administration domains, for example.}, a malicious 
node could perform MAC spoofing and exploit the relay mechanism in order to have access 
to the secure network. In this case, securing the multicast group 
by performing authentication and encryption at the multicast group level could 
prevent this kind of attacks although it may require infrastructure support. 

In conclusion, we can consider the three countermeasures introduced at the beginning of this 
section, to be more than adequate in avoiding exploitation of the relaying mechanism.

\section{Streaming Media Support} \label{sip}
SIP can be used, among other things, to update new and ongoing 
media sessions. In particular, the IP address of one or more 
of the participants to the media session can be updated. 
In general, after an MN performs a L3 handoff,
a media session update is required to inform the various 
parties about the MN's new IP address\,\cite{SIPmob}.

If the CN does not support cooperation, 
the relay mechanism as described in Section \ref{auth-relay} does not work and 
the CN keeps sending packets to the MN's old IP address, 
being unaware of the relay process. 
This is the 
case for example, of an MN establishing a streaming video session 
with a stream media server.
In this particular case, assuming that the media server supports SIP, 
a SIP session update is performed to inform the media server that the 
MN's IP address has changed. The MN sends a re-INVITE to the media server
updating its IP address to the RN's IP address. In this way, the media server 
starts sending packets to the RN and relay can take place as described earlier.

Once the 
relaying is over, if the MN's authentication was successful, the 
MN sends a second re-INVITE including its new IP address, otherwise, 
once the timeout for relaying expires, the relaying process stops 
and the RN terminates the media session with the media server. 

SIP and media session updates will be discussed further in Section \ref{SIPmob}.

\section{Bandwidth and Energy Usage} \label{energy}
By sharing information, the MNs 
in the network do not have to perform individual 
tasks such as scanning, which would normally consume a considerable 
amount of bandwidth and energy. This means that sharing
data among MNs is usually more energy and bandwidth efficient 
than having each MN perform the correspondent individual task. We 
discuss the impact of CR on energy and bandwidth below.

In CR, bandwidth usage and energy expended are mainly determined by the number of 
multicast packets that each client has to send for acquiring the 
information it needs.  The number of multicast packets is directly 
proportional to the number of clients supporting the protocol that 
are present in the network. In general, more clients introduce more requests 
and more responses. However, having more clients that support the 
protocol ensures that each client can collect more information 
with each request, which means that overall each client will
need to send fewer packets. 
Furthermore, by having the 
INFORESP frames as multicast frames, many MNs 
will benefit from each response and not just the MN that sent 
the request. This will minimize the number of packets exchanged, 
in particular the number of INFOREQ sent.

To summarize, with increasing number of clients, suppression of 
multicast takes place, so the number of packets sent remains constant.

In general, sending a few long packets is more efficient than 
sending many short ones. 
As explained in Section \ref{L2-roam}, for each AP the information included 
in an INFOREQ or INFORESP packet is a cache entry (see Fig. \ref{cacheStruct}), 
that is, a triple \{BSSID, Channel, Subnet ID\} 
for a total size of 6+4+4 = 14 bytes. Considering that
an MTU size is 1500 bytes, that each cache entry takes about 14 bytes, and 
that IP and UDP headers take together a total of 28 bytes, 
each INFOREQ and INFORESP packet can carry information about 
no more than 105 APs for a maximum of 1472 bytes. 

In \cite{measure} Henderson et al., analyze the behavior of 
wireless users in a campus-wide wireless network over a period 
of seventeen weeks. They found that:

\begin{itemize}{}
\item Users spend almost all of their time at their home 
location. The home location is defined as the AP where they 
spend most of the time and all the APs within 50 meters of 
this one.
\item The median number of APs visited by a user is 12, but 
the median differs for each device type, with 17 for laptops, 
9 for PDAs and 61 for VoIP devices such as VoIP phones.
\end{itemize}

This shows that most devices will spend most of their time 
at their home location, which means that they will mostly 
deal with a small number of APs. However, even if we 
consider the median number of APs that clients use throughout 
the trace period of seventeen weeks, we can see that when 
using laptops and PDAs each MN would have to know about 
the nearest 9-17 APs. For VoIP devices that are always on, 
the median number of APs throughout the trace period is 
61. 
In our implementation each INFOREQ and INFORESP packet carries information 
about 105 APs at most. Regardless of the type of device used, 
information about the next possible APs fits in a 
single INFOREQ or INFORESP frame. 
Also, once the cache has been filled, 
spending most of the time at the home location means 
that no new data is needed for most of the time, thus minimizing 
the number of INFOREQ and INFORESP packets exchanged.

The relay mechanism introduced in Section \ref{auth} for cooperative
authentication introduces some 
bandwidth overhead. This is because for each packet that
has to be sent by the MN to the CN and vice-versa, the packet occupies the
medium twice; once when being transmitted between MN and RN and 
once when being transmitted between RN and AP.
This however, happens only for the few seconds needed by the MN 
to authenticate. Furthermore, both of the
links MN-RN and RN-AP are maximum bit-rate links, so the \textit{time on air}
for each data packet is small.

\section{Experiments} \label{exper}
In the present section we describe implementation details
and measurement results for CR.

\subsection{Environment}
All the experiments were conducted at Columbia University on 
the 7th floor of the Schapiro building. We used four IBM Thinkpad 
laptops: three IBM T42 laptops using Intel 
Centrino Mobile technology with a 1.7 GHz Pentium processor and 1GB RAM and 
one IBM laptop with an 800 MHz Pentium III processor and 384 MB RAM.
Linux kernel version 2.4.20 was installed on all the laptops.
All the laptops were equipped with a Linksys PCMCIA Prism2 wireless card. 
Two of them were used as wireless sniffers, one of them
was used as roaming client and one was used as ``helper'' to the roaming
client, that is, it replied to INFOREQ frames and behaved as an A-MN.
For cooperative authentication the A-MN was also used as RN. Two
Dell Dimension 2400 desktops were used, one as CN and the other as
RADIUS server\,\cite{rfcRadius}.
The APs used for the experiments were a Cisco AP1231G which is an 
enterprise AP and a Netgear WG602 which is a SOHO/home AP.

\begin{figure}
\centering
\rotatebox{270}{\includegraphics[width=2.0in]{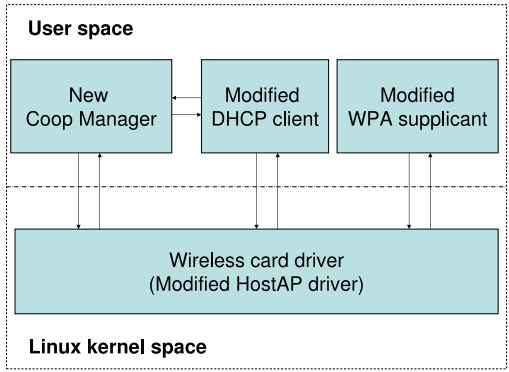}}
\caption{Implementation module interaction}
\label{arch}
\end{figure}

\begin{figure}
\centering
\rotatebox{270}{\includegraphics[width=2.0in]{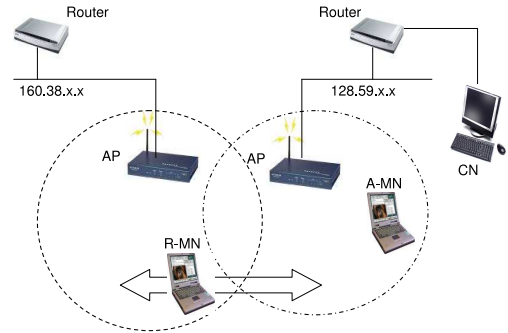}}
\caption{L3 handoff environment}
\label{L3env}
\end{figure}

\subsection{Implementation Details}
In order to implement the cooperation protocol we modified
the wireless card driver and the DHCP client. Furthermore,
a cooperation manager was also created in order to preserve state 
information and coordinate wireless driver and DHCP client.
For cooperative authentication, the WPA supplicant was also slightly 
modified to allow relay of unencrypted frames.
The HostAP\,\cite{hostap} wireless driver, an open-source WPA supplicant\,\cite{supplicant}, 
and the ISC DHCP client\,\cite{dhcp} were chosen for the implementation.
The different modules involved and their interaction is depicted 
in Fig. \ref{arch}.
A UDP packet generator was also used to generate small packets with a 
packetization interval of 20\,ms in order to simulate voice traffic.
For the authentication measurements, we used FreeRADIUS\,\cite{radius} as 
RADIUS server.

\subsection{Experimental Setup}
For the experiments we used the Columbia University 802.11b wireless network which
is organized as one single subnet. In order to 
test L3 handoff, we introduced another AP connected to a different subnet 
(Fig. \ref{L3env}). The two APs operated on two different non-overlapping channels.

The experiments were conducted by moving the roaming client between two 
APs belonging to different subnets, thus having the client perform L2 and 
L3 handoffs in either direction.

Packet exchanges and handoff
events were recorded using the two wireless sniffers (kismet\,\cite{kismet}), one per channel.
The trace files generated by the wireless sniffer were later analyzed using 
Ethereal\,\cite{ether}.

\begin{figure*}
\centering
\rotatebox{270}{\includegraphics[height=6.0in]{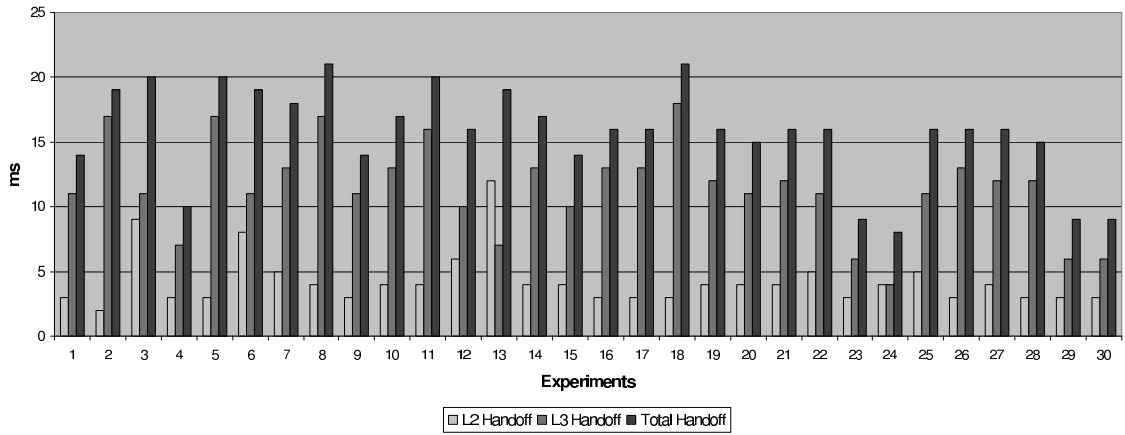}}
\caption{Measured L2 and L3 handoff time with CR in an open network}
\label{htime}
\end{figure*}

In the experimental set-up we do not consider 
a large\footnote{Other MNs were present in the Columbia wireless network during the experiments.} presence 
of other MNs under the same AP since air-link congestion is not relevant to the handoff measurements.
Delays due to collisions, backoff, propagation delay and AP queuing delay are irrelevant 
since they usually are on the order of micro-seconds under normal conditions. However, even if we consider these 
delays to be very high because of a high level of congestion, the MN should worry about not being able to 
make or continue a call as the AP has reached its maximum capacity. Handoff delay would, at this point, become a 
second order problem. Furthermore, in this last scenario, the MN should avoid to do handoff to a very 
congested AP in the first place as part of a good handoff policy (see Section \ref{loadb}).
Updating information at the Home Agent or SIP Registrar is trivial and does not have the same stringent 
delay requirements that mid-call mobility has, therefore it will not be considered. 

\subsection{Results} \label{res}
In this section we show the results
obtained in our experiments. In Section \ref{roamres},
we consider an open network with no authentication in order to 
show the gain of CR in an open network. In Section \ref{authres}, 
authentication is added and, in particular, we consider
a wireless network with IEEE 802.11i enabled. 

We define L2 handoff time as scanning time + open authentication and 
association time + IEEE 802.11i authentication time. 
The last contribution to the L2 handoff time is not present in open networks. 
Similarly, we define the L3 handoff time as subnet discovery time + IP address 
acquisition time.

In the following experiments we show the drastic improvement achieved by CR in terms of 
handoff time. At L2 such an improvement is possible because, as 
we have explained in Section \ref{L2-roam-ovrvw}, MNs build a cache of 
neighbor APs so that scanning for new APs is not required 
and the delay introduced by the scanning procedure during the L2 
handoff is removed. Furthermore, by using relays (see Section \ref{auth}), an MN can send and receive data packets 
\textit{during} the authentication process, thus eliminating the 802.11i authentication delay.
At L3, MNs cache information about which AP belongs to which subnet, hence immediately detecting 
a change in subnet by comparing the subnet IDs of the old and new APs. This provides a way to 
detect a subnet change and at the same time makes the subnet discovery delay insignificant. 
Furthermore, with CR, the IP address acquisition delay is completely removed since each node can acquire a new 
IP address for the new subnet while still in the old subnet (see Section \ref{L3-roam}).

It is 
important to notice that in current networks\footnote{Within the IETF, the DNA working 
group is standardizing the detection of network attachments for IPv6 networks only\,\cite{draftDNA}.} there is no 
standard way to detect a change in subnet in a timely manner\footnote{Router advertisements are typically broadcast 
only every few minutes.}. 
Recently, DNA for IPv4 (DNAv4)\,\cite{dnav4} was standardized 
by the DHC working group within the IETF in order to detect a subnet change in IPv4 networks. This mechanism, 
however, works only for previously visited subnets for which the MN still has a valid IP address and 
can take up to hundreds of milliseconds to complete. Furthermore, if L2 authentication is used, a change 
in subnet can be detected only after the authentication process completes successfully.
Because of this, in the handoff time measurements for the standard IEEE 802.11 handoff procedure, the delay introduced by subnet change 
discovery is not considered.

\begin{table}
\centering
\caption{Performance overview for CR (average)}
\begin{tabular}{|l|r|} \hline
IP\_REQ - IP\_RESP & 867.0 ms\\ \hline
L2 handoff & 4.2 ms\\ \hline
L3 handoff & 11.4 ms\\ \hline
Total handoff & 15.6 ms\\ \hline
Packet loss & 1.3 packets\\ \hline
\end{tabular}
\label{htable}
\end{table}

\begin{figure}
\centering
\rotatebox{270}{\includegraphics[width=2.3in]{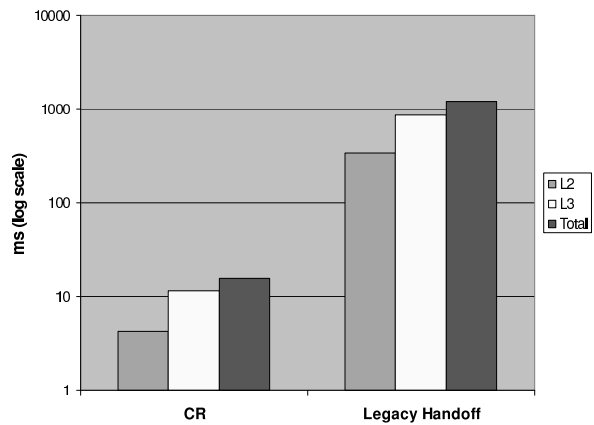}}
\caption{Average handoff time for CR and IEEE 802.11b in an open network}
\label{hcomp}
\end{figure}

To summarize, in theory by using CR the only contribution to the L2 handoff time is given by open authentication 
and association 
and there is no contribution to the L3 handoff time whatsoever, that is, the L3 handoff time is zero. In practice, this 
is not exactly true. Some other sources of delay have to be taken into consideration as we show in more 
detail in Section \ref{variance}.

\subsubsection{L2 and L3 Roaming} \label{roamres}
We show the handoff time 
when an MN is performing a L2 and L3 handoff without
any form of authentication, that is, the MN is moving in an open
network.
In such a scenario, before the L2 handoff occurs, the MN tries to 
build its L2 cache if it has not already done so. Furthermore, the MN also searches
for any available A-MN that might help it in acquiring an IP address for
the new subnet. 
The scenario is the same as the one depicted in Fig. \ref{L3env}.

Fig. \ref{htime} shows the handoff time when CR is used. In particular, 
we show the L2, L3 and total L2+L3 handoff times over 30 handoffs. 
As we can see, the total L2+L3
handoff time has a maximum value of 21\,ms in experiment 18.
Also, we can see how, even though the L3 handoff time is higher on average than
the corresponding L2 handoff time, there are situations where these two become comparable.
For example, we can see in experiment 24 how the L2 and L3 handoff times are equal
and in experiment 13 how the L2 handoff time exceeds the corresponding L3 handoff time.
The main causes for this variance will be presented in Section \ref{variance}.

\begin{figure}
\centering
\rotatebox{270}{\includegraphics[width=2.2in]{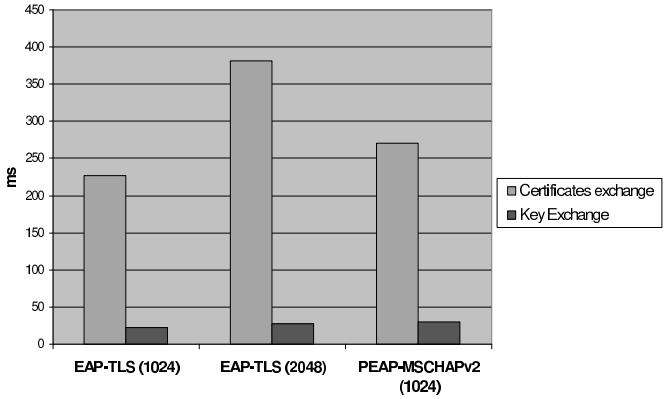}}
\caption{Authentication delay in IEEE 802.11i}
\label{auth_delay}
\end{figure}

Fig. \ref{htime} and Table \ref{htable} show how, on average, with CR the total L2+L3 handoff time is 
less than 16\,ms, which is less than half of the 50\,ms requirement for assuring 
a seamless handoff when real-time traffic is present. 

Table \ref{htable} shows the average values of IP address acquisition time, 
handoff time, and packet loss during the handoff process.
The time between IP\_REQ and IP\_RESP is the time needed by the A-MN to acquire a new IP
address for the R-MN. This time can give a good approximation of the L3 handoff time that we would
have without cooperation. As we can see, with cooperation we reduce the L3 handoff time
to about 1.5\% of what we would have without cooperation.
Table \ref{htable} also shows that the packet loss experienced during a L2+L3 handoff is 
negligible when using CR.

Fig. \ref{hcomp} shows the average delay over 30 handoffs of L2, 
L3 and L2+L3 handoff times for CR and for the legacy 802.11 handoff mechanism.  
The total L2+L3 handoff time is less than 16\,ms for CR while it is about 1210\,ms for the 
legacy 802.11 handoff mechanism. CR has reduced the total handoff time to 1.3\% of the handoff time introduced 
by the standard 802.11 handoff procedure.

\subsubsection{L2 and L3 Roaming with Authentication} \label {authres}
Here we show the handoff time when IEEE 802.11i is used together with EAP-TLS 
and PEAP/MSCHAPv2.

\begin{figure}
\centering
\rotatebox{270}{\includegraphics[width=2.2in]{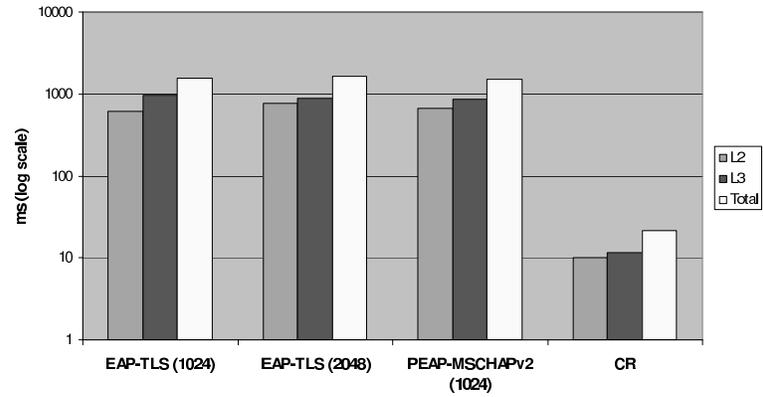}}
\caption{Handoff time in IEEE 802.11i networks}
\label{auth_time}
\end{figure}

\begin{figure*}
\centering
\rotatebox{270}{\includegraphics[height=6.0in]{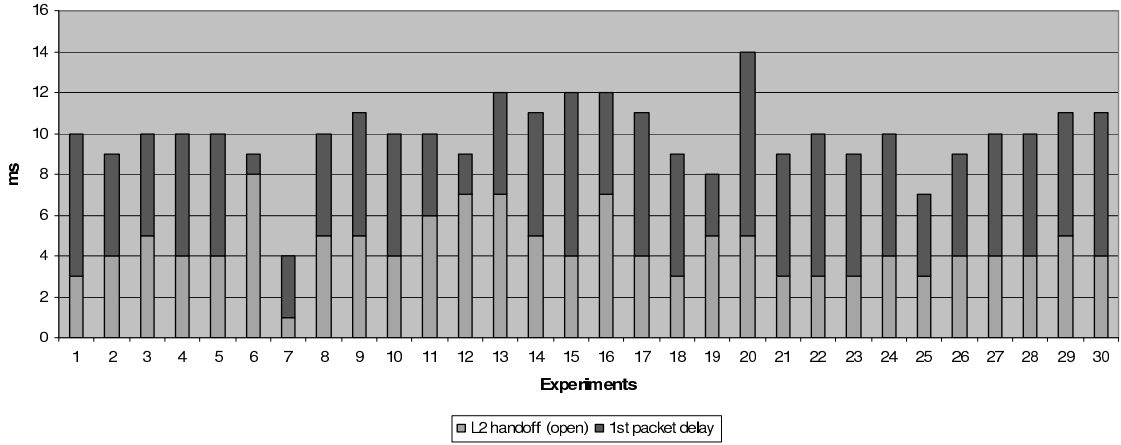}}
\caption{CR L2 handoff time in IEEE 802.11i networks}
\label{auth_relay}
\end{figure*}
Fig. \ref{auth_delay} shows the average over 30 handoffs of the delay introduced in a L2 handoff by 
the certificate/credentials exchange and the session key exchange. Different key 
lengths are also considered for the generation of the certificates\footnote{The length of certificates affects
the handoff time much more than the length of session keys.}. 
As expected,
the exchange of certificates takes most of the time. This is the reason why mechanisms 
such as fast-reconnect\,\cite{rfcEAP, draftEAP} improve L2 handoff times considerably, although 
still on the order of hundreds of milliseconds.

Generally speaking, any authentication mechanism can be used together with CR.
Fig. \ref{auth_time} shows the average over 35 handoffs of the total L2, L3 
and L2+L3 handoff times.
In particular, we show the handoff time for EAP-TLS with 1024 and 2048 bits key, 
PEAP/MSCHAPv2 with 1024 bits key and CR. The 
average L2+L3 handoff times are
respectively 1580\,ms, 1669\,ms, 1531\,ms and 21\,ms. 
By using CR, we achieve a drastic improvement in the total handoff time. As we can see, 
CR reduces the handoff time to 1.4\% or less of the handoff time introduced by the standard 802.11 mechanism.
This significant improvement is possible because at L2 with CR we bypass the whole authentication handshake 
by relaying packets. At L3 we are able to detect a change in subnet in a timely manner and acquire 
a new IP address for the new subnet while still in the old subnet. 

Fig. \ref{auth_relay} shows in more detail the two 
main contributions to the L2 handoff time when a relay is used. We can see 
that, on average, the time needed for the first data packet to be transmitted 
after the handoff takes more than half of the total L2 handoff time. Here, with 
data packet we are referring to a packet sent by our UDP packet generator.
By analyzing the wireless traces collected in our experiments, we found that
the first data packet after the handoff is not transmitted immediately after
the L2 handoff completes because the wireless driver needs to start 
the handshake for the authentication process. This means that the driver already 
has a few packets in the transmission queue that are waiting to be 
transmitted when our data packet enters the transmission queue. This, 
however, concerns only the first packet to be transmitted after the L2 handoff
completes successfully. All subsequent data packets will not encounter
any additional delay due to relay.

\subsubsection{Measurement Variance} \label{variance}
We have encountered a 
high variance in the L2 handoff time. In particular, most of the delay is
between the authentication request and authentication response, before the association request. 
Within all the measurements taken, such behaviour appeared to be particularly prominent when moving 
from the Columbia AP to the Netgear AP. This behavior, together with the results shown by Mishra et al.
in \cite{L2analysis}, have lead us to the conclusion that such a variance 
is caused by the cheap hardware used in the low-end Netgear AP.

At L3, ideally, the handoff time should be zero as we acquire all the
required L3 information while still in the old subnet. The L3 handoff time shown in Fig. \ref{htime}
can be roughly divided in two main components: \textit{signaling delay} and \textit{polling delay}.
The signaling delay is due to various signaling messages exchanged among the 
different entities involved in setting up
the new L3 information in the kernel (wireless driver and DHCP client);
the polling delay is introduced by the polling of variables in between received-signal-strength
samples\footnote{Received-signal-strength values are measured by the wireless card driver.}, 
done in order to start the L3 handoff process in a timely manner with respect to the
L2 handoff process.

These two delays are both implementation dependent
and can be reduced by further optimizing the implementation. 

\section{Application Layer Mobility} \label{SIPmob}
We suggest a method for achieving seamless handoffs at the 
application layer using SIP and CR. Implementation 
and analysis of the proposed approach are reserved for future work.

Generally speaking, there are two main problems with application layer mobility. One is 
that the SIP handshake (re-INVITE $\Rightarrow$ 200 OK $\Rightarrow$ ACK) takes a few hundred milliseconds to 
complete, exceeding the requirements of seamless handoff for real-time media.
The second is that we do not know \textit{a priori} in which direction the user is going to move.

In order to solve these two problems, we have to define a mechanism that allows the MN 
to start the application layer handoff before the L2 handoff and to do it so that the MN does not 
move to the wrong AP or subnet after updating the SIP session. Furthermore, the new mechanism also has to work in
the event of the MN deciding not to perform the L2 handoff at all after performing the SIP session update, that is,
after updating the SIP session with the new IP address.

\begin{figure}
\centering
\begin{tabular}{|l|} \hline
v=0\\
o=Laura 289083124 289083124 IN IP4 five.example.com\\
t=0 0\\
c=IN IP4 131.160.1.112\\
a=group:FID 1 2 3\\
m=audio 30000 RTP/AVP 0\\
a=mid:1\\
m=audio 30002 RTP/AVP 8\\
a=mid:2\\
m=audio 20000 RTP/AVP 0 8\\
c=IN IP4 131.160.1.111\\
a=mid:3\\ \hline
\end{tabular}
\caption{RFC3388, SDP format}
\label{SDPtable}
\end{figure}

The SIP mobility mechanism\,\cite{SIPmob} and CR can be combined.
In particular, we consider an extension of the relay mechanism discussed in Section \ref{auth-relay}.
Let us assume that the MN performing the handoff has already acquired all the necessary L2 and L3 information
as described in Sections \ref{L2-roam}, \ref{L3-roam} and \ref{auth}. This means that the MN has a list of 
possible RNs and IP addresses to use after the L2 handoff, one for each of 
the various subnets it could move to next. At this point, before performing any L2 handoff, the MN needs to 
update its multimedia session. The up-link traffic does not cause particular problems as the MN already
has a new IP address to use and can start sending packets via the RN right after the L2 handoff. The 
down-link traffic is more problematic since the CN will continue sending packets to the MN's old IP address
as it is not aware of the change in the MN's IP address until the session has been updated.

The basic idea is to update the session so that the same media stream is sent, at the same time, to the 
MN and to all the RNs in the list previously built by the MN.
In this way, regardless of which subnet/AP the MN will move to, the corresponding RN will 
be able to relay packets to it. If the MN does not change AP at all, nothing is lost as the MN
is still receiving packets from the CN.
After the MN has performed the L2 handoff and has connected to one of the RNs, it may send
a second re-INVITE via the RN so that the CN sends packets to the current RN only, without involving the other
RNs any longer. Once the authentication process successfully completes, communication
via the AP can resume. At this point, one last session update is required so that the CN can send packets
directly to the MN without any RN in between.

In order to send multiple copies of the same media stream to different nodes, that is, to the MN performing the
handoff and its RNs, the MN can send to the CN a re-INVITE with an SDP format as described in RFC 3388\,\cite{rfc3388} 
and shown in Figure \ref{SDPtable}. In this particular format, multiple \textit{m} lines are 
present with multiple \textit{c} lines and grouped together by using the same Flow Identification (FID).
A station receiving a re-INVITE with an SDP part as shown in Figure \ref{SDPtable} sends an audio stream 
to a client with IP address 131.160.1.112 on port 30000 (if the PCM $\mu$-law codec is used) and to a client with IP address 131.160.1.111 on 
port 20000. In order for the same media 
stream to be sent to different clients at the same time, all the clients have to support the same codec\,\cite{rfc3388}. In our case, we have to 
remember that RNs relay traffic to MNs, they do not play such traffic. Because of this, we can safely say that 
each RN supports any codec during the relay process, hence a copy of the media stream can always be sent 
to an RN by using the SDP format described in \cite{rfc3388}.

It is worthwhile to notice that in the session update procedure described above, no buffering 
is necessary. As we have explained in Section \ref{res} and shown in Table \ref{htable}, 
the L2+L3 handoff time is on the order of 16\,ms for open networks, which is less than the packetization 
interval for typical VoIP traffic. When authentication is used (see Figure \ref{auth_time}), the total L2+L3 handoff time is on 
the order of 21\,ms. In both cases packet loss is negligible, hence 
making any buffering of packets unnecessary.

\section{Load Balancing} \label{loadb}
CR can also play a role in AP load balancing.
Today, there are many problems with the way MNs select the AP to connect
to. The AP is selected according to the link signal strength and
SNR levels while other factors such as effective throughput, number of retries,
number of collisions, packet loss, bit-rate or BER are not taken into account.
This can cause an MN to connect to an AP with the
best SNR but low throughput, high number of collisions and packet loss
because that AP is highly congested. If the MN disassociates or the AP deauthenticates it, 
the MN looks for a new candidate AP. Unfortunately, with a very high probability, the MN 
will pick the same AP because its link signal strength and SNR are still the ``best'' 
available. The information regarding 
the congestion of the AP is completely ignored and this bad behavior keeps repeating itself.
This behavior can create situations where users end up connecting all to 
the ``best'' AP creating the scenario depicted earlier and at the same time leaving 
other APs under-utilized \cite{Jardosh05, ietf65}. 

CR can be very helpful in such a context. In particular, we can imagine a situation where
an MN wants to gather statistics about the APs that it might move to next, that is, the APs
that are present in its cache. In order to do so, the MN can ask other nodes to
send statistics about those APs. Each node can collect different kind of statistics, such
as available throughput, bit-rate, packet loss, retry rate. 
Once these statistics have been gathered, they
can be sent to the MN that has requested them. The MN, at this point has a clear picture of which
APs are more congested and which others can support the required QoS, therefore making a smarter handoff
decision. By using this approach we can achieve an even distribution of traffic flows among neighboring
APs.

The details of this mechanism are reserved for future study but can be easily
derived from the procedures earlier introduced for achieving fast L2 and L3 handoffs.

\section{An Alternative to Multicast} \label{alternative}
Using IP multicast packets can become inefficient in 
highly congested environments with a dense distribution of 
MNs. In such environments a good alternative to multicast 
can be represented by ad-hoc networks. 
Switching back-and-forth between infrastructure mode and ad-hoc mode has already 
been used by MNs in order to share information for fault diagnosis\,\cite{wifi-profiler}. 
As we pointed out in Section \ref{relay-proc}, continuously switching between ad-hoc and infrastructure 
mode introduces synchronization problems and channel switching delays, making this approach unusable for 
real-time traffic. However, even if non-real-time traffic is present, synchronization problems could still arise 
when switching to ad-hoc mode 
while having an alive TCP connection on the infrastructure network, for example. Spending a longer time in 
ad-hoc mode might cause the TCP connection to time-out; on the other hand waiting too 
long in infrastructure mode might cause loss of data in the ad-hoc network. 
 
In CR, MNs can exchange L2 and L3 information contained in their cache by 
using the mechanism used for relay as described in Section \ref{relay-proc}. Following this approach, 
MNs can directly exchange information with each other without involving the AP and 
without having to switch their operating mode to ad-hoc.
In particular, an MN can send broadcast and unicast packets such as INFOREQ and INFORESP with 
the \textit{To DS} and \textit{From DS} fields set to zero (see Section \ref{relay-proc}). Because 
of this, only the MNs in the radio coverage of the first MN will be able to receive such packets. The 
AP will drop these packets since the \textit{To DS} field is not set. 

Ad-hoc multi-hop routing can also 
be used when needed. This may be helpful, for example, in the case of R-MNs acquiring a new IP address 
for a new subnet while still in the old subnet (see Section \ref{L3-roam}), when current AP and new AP 
use two different channels. In such a case, a third node on the same channel than the R-MN, could route packets between 
the R-MN and the A-MN by switching between the two channels of the two APs, thus leaving R-MN and A-MN operations 
unaffected. In this case we would not have synchronization problems since the node, switching between the 
two channels, would have to switch only twice. Once after receiving the IP\_REQ packet from the R-MN in 
order to send it to the A-MN, and a second time after receiving the IP\_RESP from the A-MN in order to 
send it to the R-MN.

An ad-hoc based approach, such as the relay mechanism presented in Section \ref{relay-proc}, does not 
require any support on the infrastructure and 
it represents an effective solution in congested and densely populated environments. 
On the other hand, ad-hoc communication between MNs would not work 
very well in networks with a small population of MNs, where each MN 
might be able to see only a very small number of other MNs at any given time.

MNs with two wireless cards could use one card to connect to the ad-hoc network 
and share information with other MNs, while having the other card 
connected to the AP. The two cards could also operate on two different access technologies such as 
cellular and 802.11.

If it is possible to introduce some changes in the infrastructure, we can 
minimize the use of multicast packets by using the SIP presence model\,\cite{rfcpresence}. 
In such a model we 
introduce a new presence service in which each subnet is a 
presentity. Each subnet has a contact list of all the 
A-MNs available in that subnet for example, so that the presence information 
is represented by the available A-MNs in the subnet. When 
an R-MN subscribes to this service, it receives presence 
information about the new subnet, namely its contacts which 
are the available A-MNs in that subnet.

This approach could be more efficient in scenarios with a small 
number of users supporting CR. On the other hand, it 
would require changes in the infrastructure by introducing additional 
network elements. The presence and ad-hoc approaches are reserved for future study.

\section{Conclusions and Future Work} \label{concl}
In this paper we have defined the Cooperative Roaming protocol. 
Such a protocol allows MNs to perform L2 and L3 handoffs seamlessly, with
an average total L2+L3 handoff time of about 16\,ms in an open network and of about 21\,ms 
in an IEEE 802.11i network without requiring any changes 
to either the protocol or the infrastructure. Each of these values is less than half of the 
50\,ms requirement for real-time applications such as VoIP to achieve seamless
handoffs. Furthermore, we are able
to provide such a fast handoff regardless of the particular authentication
mechanisms used while still preserving security and privacy.

MN cooperating has many advantages and does not 
introduce any significant disadvantage as in the worst case scenario MNs can 
rely on the standard IEEE 802.11 mechanisms achieving performances similar to a 
scenario with no cooperation.

Node cooperation can be useful in many other applications:

\begin {itemize}
\item In a multi-administrative-domain environment CR can help in discovering which 
APs are available for which domain. In this way an MN might decide to 
go to one AP/domain  rather than some other AP/domain according to 
roaming agreements, billing, etc.
\item In Section \ref{loadb} we have shown how CR can be used for load balancing. Following 
a very similar approach but using other metrics such as collision rate and available bandwidth, 
CR can also be used for admission control and call admission control.
\item CR can help in propagating information about service availability. In 
particular, an MN might decide to perform a handoff to one particular AP 
because of the services that are available at that AP. A service might be a 
particular type of encryption, authentication, minimum guaranteed bit rate and 
available bandwidth or the availability of other types of networks such as 
Bluetooth, UWB and 3G cellular networks, for example.
\item CR provides advantages also in terms of adaptation to changes in the 
network topology. In particular, when an MN finds some stale entries in its 
cache, it can update its cache and communicate such changes to the other MNs. 
This applies also to virtual changes of the network topology (i.e. changes 
in the APs power levels) which might become more common with the deployment of 
IEEE 802.11h equipment.
\item CR can also be used by MNs to negotiate and adjust their transmission power
levels so to achieve a minimum level of interference.
\item In \cite{syncscan} Ramani et al. describe a passive scanning algorithm
according to which an MN knows the exact moment when
a particular AP will send its beacon frame. In this way the MN collects the 
statistics for the various APs using passive scanning but without having to 
wait for the whole beacon interval on each channel. This algorithm, however, 
requires all the APs in the network to be synchronized. 
By using a cooperative approach, we can have the various MNs sharing information
about the beacon intervals of their APs. In this way, we only need to have
the MNs synchronized amongst themselves (e.g., via NTP) without any 
synchronization required on the network side.
\item Interaction between nodes in an infrastructure network and nodes in an 
ad-hoc/mesh network.
\begin {enumerate}
\item An MN in ad-hoc mode can send information about its ad-hoc network. In 
this way MNs of the infrastructure network can decide if it is convenient for 
them to switch to the ad-hoc network (this would also free resources on the 
infrastructure network). 
This, for example, can happen if there is a lack of coverage or if there is high 
congestion in the infrastructure network. Also, an MN might switch to an ad-hoc 
network if it has to recover some data available in the ad-hoc network itself 
(i.e. sensor networks).
\item If two parties are close to each other, they can decide to switch to 
the ad-hoc network discovered earlier and talk to each other without any 
infrastructure support. They might also create an ad-hoc network on their own 
using a default channel, if no other ad-hoc network is available.
\end{enumerate}
\end{itemize}

As future work, we will look in more detail at application layer 
mobility, load balancing and call admission control. We will investigate the possibility of having some 
network elements such as APs support A-MN and RN functionalities; this would be 
useful in scenarios where only few MNs support CR. Finally, we will look 
at how IEEE 802.21\,\cite{80221_draft} could integrate and extend CR.

\section*{Acknowledgments}
This work was supported by grant of FirstHand Technologies. Equipment was 
partially funded through grant CISE 02-02063 of the National Science Foundation.

\bibliographystyle{IEEEtran}
\bibliography{IEEEabrv,Cooperative_Roaming}

\end{document}